\documentclass[11pt]{cernrep}
\usepackage{graphicx}
\usepackage{here}
\begin{document}
 \title{Recovering Lorentz Invariance of DLCQ
\footnote{Talk given at Light-Cone Workshop ``Hadrons and Beyond''
(LC03), Durham, 2003.}}
\author{ Masa-aki Taniguchi, Shozo Uehara, Satoshi Yamada
	 and Koichi Yamawaki}
\institute{Department of Physics, Nagoya University, Nagoya, Japan}
\maketitle
\vspace{-40mm}
\begin{flushright}
       DPNU-03-30 \\ hep-th/0309240
\end{flushright}
\vspace{20mm}
\begin{abstract}
We propose a way to recover Lorentz invariance of the perturbative
S matrix in the Discrete Light-Cone Quantization (DLCQ) in the
continuum limit without spoiling the trivial vacuum.
\end{abstract}

\section{INTRODUCTION}
The Discrete Light-Cone Quantization (DLCQ) was first introduced by
Maskawa and Yamawaki (MY) \cite{M&Y} in 1976 based on the canonical
quantization for the constrained system due to Dirac and was also
considered by Casher \cite{Casher} slightly later in 1976 and by Pauli
and Brodsky \cite{P&B} in 1985 in different contexts.
In the original paper of MY \cite{M&Y}, the light-like coordinate
$x^-$ was compactified $-L \leq x^- \leq L$ with periodic boundary
condition thus discretizing the conjugate momentum, $p^+ = (n\pi)/L
\quad (n=0,\pm 1,\pm 2,...)$, in order to isolate the zero mode
($n=0$), $\phi_0\equiv \frac{1}{2L}\int_{-L}^{L}dx^-\phi $, from the
non-zero mode, $\varphi \equiv \phi-\phi_0$.

The most important finding of MY \cite{M&Y} is the discovery of the
{\it Zero-Mode Constraint (ZMC)}:
\begin{equation}
\Phi_3 \equiv \frac{1}{2L}\int^{L}_{-L}d x^-
	\left[(m^2- \partial_{\bot}^2)\phi
	+\frac{\partial V(\phi)}{\partial \phi}\right]
      =\left[\frac{\partial \mathcal{H}}{\partial \phi}\right]_0
      =0 \, \label{ZMC0}
\end{equation}
in the 4-dimensional scalar theory with mass $m$ and self-interaction
$V(\phi)$, where $\mathcal{H}= \frac{1}{2}\phi(m^2 -
\partial_{\bot}^2)\phi +V(\phi)$ is the light-cone Hamiltonian density
and $[\mathcal{O}]_0 \equiv \frac{1}{2L}\int_{-L}^{L}dx^- \mathcal{O}$
is the zero mode of the operator $\mathcal{O}$.
 Thanks to ZMC, MY succeeded to compute the {\it well-defined Dirac
bracket} at $x^+=y^+$:
\begin{equation}
 \{\phi(\vec{x}), \phi(\vec{y})\}_D \equiv
  \{\phi(\vec{x}),\phi(\vec{y})\}- \int d\vec{z} d\vec{z}^{\prime}
    \{\phi(\vec{x}), \Phi_i(\vec{z})\}
    C^{-1}_{ij}(\vec{z},\vec{z}^\prime)
    \{\Phi_j(\vec{z}^\prime),\phi(\vec{y})\}
\label{Dbracket}
\end{equation}
where for convenience the primary second-class constraint
$\Phi \equiv \pi -\partial_{-}\phi$ may be divided into the non-zero
mode $\Phi_1 \equiv \pi_{\varphi} - \partial_{-}\varphi$ and the zero
mode $\Phi_2 \equiv [\Phi]_0= \pi_0$, and $C^{-1}$ is the inverse of
$C_{ij}(\vec{x},\vec{y}) \equiv \{\Phi_i(\vec{x}), \Phi_j(\vec{y})\}$,
with $\vec{x}\equiv (x^-, x^{\bot})$ and $(i,j)=1,2,3$.
Without ZMC, $\Phi_3$, which yields $C_{i,3}=-C_{3,i} \ne0 \quad
(i=1,2)$, we would have trouble to get the inverse $C^{-1}$
and hence the Dirac bracket, since $C_{1,2}=C_{2,1}=C_{2,2}=0$.
Thus the ZMC is vital to a well-defined canonical light-cone
commutator $[\phi(x),\phi(y)]_{x^+=y^+}$ which is given as $i\hbar$
times the Dirac bracket (up to operator ordering).

{}From the ZMC (\ref{ZMC0}) and the Dirac bracket (\ref{Dbracket})
(hence the canonical commutator), MY obtained two important physical
consequences \cite{M&Y}:

\begin{itemize}
\item Proof of the Trivial Vacuum\\
Since the {\it translation invariance}, $[\phi,P^+]= i \partial_-\phi$
is manifest in DLCQ even for finite $L$, the vacuum is defined
independently of the dynamics (thus ``trivial'') as the lowest $p^+$
state ($p^+=0$).
This trivial vacuum proved a unique $p^+=0$ state and hence the true
vacuum, since the zero mode is {\it not an independent degree of
freedom}, written in terms of the non-zero modes,
$\phi_0=\phi_0(\varphi)$, by solving the ZMC (at least in
perturbation), and hence can be removed from the physical Fock space.

\item Violation of the Lorentz Invariance\\
Based on the canonical commutator mentioned above, the Lorentz algebra
$[\phi(x), M^{\mu\nu}]$ was explicitly computed; for the Lorentz
generators $M^{+-}, M^{{\bot}-}$ which change the quantization
plane (light-front), the Lorentz invariance is {\it violated at
operator level} due to the unwanted surface term at $x^- =\pm L$
which does not vanish even in the continuum limit ($L\to\infty$).
\end{itemize}
Thus the incompatibility between the trivial vacuum and the Lorentz
invariance was the very nature of DLCQ from the beginning back in 1976
and has been a long-standing problem. (See Ref. \cite{Yam}.)

One might consider that the trivial vacuum could be fake, namely
it could  not be reconciled with the nonperturbative phenomena such as
the confinement and the spontaneous symmetry breaking (SSB) which are
attributed to the complicated vacuum structure in the usual
quantization. However, it was demonstrated  \cite{KTY,Yam} that
{\it the operator solution of ZMC together with the trivial vacuum
can in fact describe the SSB} in four dimensions for continuous
symmetry in a way to regularize the zero mode of the Nambu-Goldstone
boson through explicit symmetry breaking which is taken to zero at the
end of all calculations. (See also footnote \ref{ft1}.)
What about the Lorentz invariance?

In this talk we shall demonstrate that
\begin{itemize}
\item The Lorentz invariance is violated {\it at S matrix level}
as well as at operator level due to {\it lack of the zero-mode loop}
in the perturbative DLCQ in the continuum limit \cite{TUYY}.
\item Lorentz invariance at perturbative S matrix level can be
recovered by {\it modifying the naive DLCQ action} into the one with
additional operator arising from the zero-mode loop in such a way that
{\it the trivial vacuum remains intact} by this modification
\cite{TUYY2}.
\end{itemize}
 
\section{DLCQ PERTURBATION}
To be definite we confine ourselves to two dimensional scalar theory
with $V(\phi)=\frac{\lambda}{4 !} \phi^4$. Extension will be discussed
in the end. The ZMC (\ref{ZMC0}) in two dimensions reads
$ m^2\phi_0 +\frac{\lambda}{3!}\left([\varphi^3]_0
   +3[\varphi^2]_0\phi_0 +\phi_0^3\right) = 0\, .\label{ZMC2}$
This is a cubic equation for $\phi_0$ once $[\varphi^3]_0$ and
$[\varphi^2]_0$ are given, and hence has three solutions (up to
operator ordering):\cite{TUYY2}
\begin{equation}
\phi_0^{(0)}= G_- - G_+, \quad \phi_0^{(\pm v)}
    = \pm \left[ e^{\frac{4\pi i}{3}} G_{\mp} - e^{\frac{2\pi i}{3}}
	G_{\pm}\right],\label{Sol}
\end{equation}
where
\begin{equation}
G_{\mp} \equiv \left(\mp \frac{1}{2}[\varphi^3]_0+\frac{1}{2}
  \sqrt{\left([\varphi^3]_0\right)^2+4\left([\varphi^2]_0
  +\frac{2m^2}{\lambda}\right)^3}\right)^{\frac{1}{3}}\, .
\end{equation}
In the case of $m^2>0$, only $\phi_0^{(0)}$ is a real solution
\footnote{In the case of double-well potential $m^2<0$, we can easily
see \cite{TUYY2} that other two solution $\phi_0^{(\pm v)}$ also
become real for weak coupling $\lambda \ll 1$ and their Taylor
expansion coincides with the perturbation around the non-zero value
$\pm v$ ($v = \sqrt{-6m^2/\lambda}$) \cite{KTY}, where $\pm v$ is
nothing but the vacuum expectation value
$\langle 0| :\phi_0^{(\pm v)}: |0 \rangle $ {\it on the trivial
vacuum} $|0 \rangle$ ($:\quad :$ is the normal ordering). This implies
that {\it the trivial vacuum is indeed the true vacuum even when SSB
takes place}.
As was emphasized by Ref. \cite{Yam}, the information of {\it SSB is
carried by the operator but not by the vacuum} which is always trivial
in contrast to the usual quantization. The three operator solutions
$\phi_0^{(0)}, \phi_0^{(\pm v)}$ are plugged into the original
Hamiltonian, yielding three {\it different effective Hamiltonians}
(without zero mode) $\mathcal{H}^{(0, \pm v)}(\varphi)$ whose
vacuum energy is $\langle 0| :\mathcal{H}^{(0,\pm v)}: |0 \rangle=0,
{}-\frac{3m^4}{2\lambda}$
{\it for the trivial vacuum} $|0\rangle$. Thus $\mathcal{H}^{(v)}$
or $\mathcal{H}^{(-v)}$, which is the {\it operator non-invariant
under $Z_2$},  yields the physical (ground-state) solution, in perfect
agreement with the result of the conventional SSB in a different
language.\label{ft1}}
whose Taylor expansion for $\lambda\ll 1$ coincides with the
perturbative solution \cite{M&R} (see Fig. 1) :
\begin{eqnarray}
\phi_0^{(0)}(x^+)&=& -i\lambda \int d^2y\,
	\Delta_{0}(x-y)\frac{\varphi^3(y)}{3!} \nonumber\\
  &&+ (-i\lambda)^2 \int d^2y_1 \int d^2y_2\,
   \Delta_{0}(x-y_1)\,\frac{\varphi^2(y_1)}{2!}\Delta_{0}(y_1-y_2)\,
	\frac{\varphi^3(y_2)}{3!}\,+ \cdots.\, ,
\end{eqnarray}
with $\int d^2 x \equiv \int d x^+ \int_{-L}^{L} d x^-$, where
$\Delta_0(x-y)=\frac{1}{2iL}{\frac{1}{m^2}\delta(x^+-y^+)}$ is the
zero-mode propagator (actually not ``propagating'' due to
instantaneous delta function $\delta(x^+-y^+)$, reflecting the fact
that the zero mode is not an independent degree of freedom).
Note that the ZMC solution produces {\it only the tree} zero-mode graph.
\begin{figure}
\begin{center}
\begin{minipage}{70mm}
\includegraphics[width=70mm]{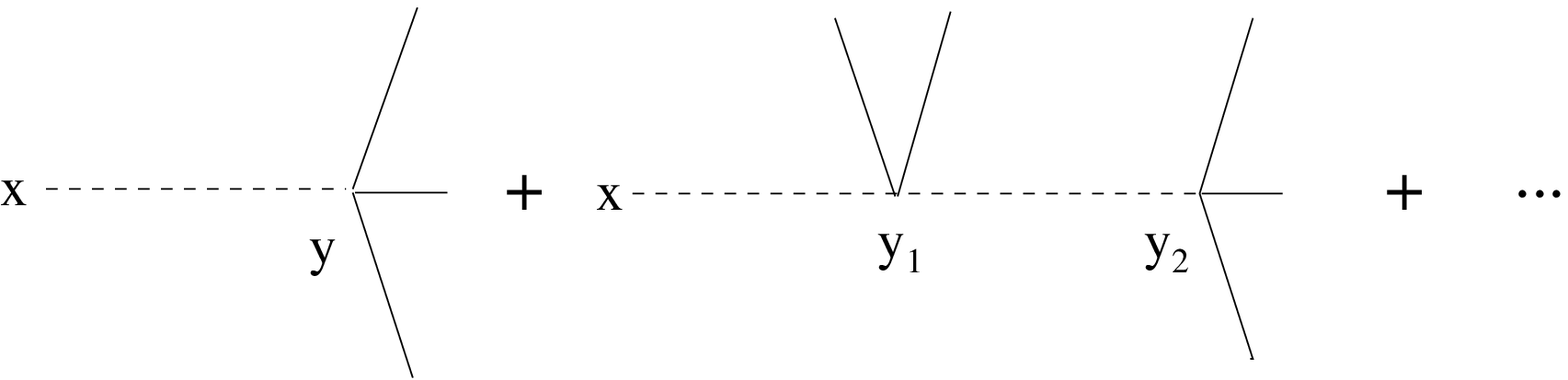}
\caption{Perturbative expansion of $\phi^{(0)}_0$. The solid line is the
non-zero mode operator $\varphi$ and the dotted line is
the zero-mode propagator $\Delta_0$.}
\end{minipage}\hspace{5mm}
\begin{minipage}{75mm}
\includegraphics[width=70mm]{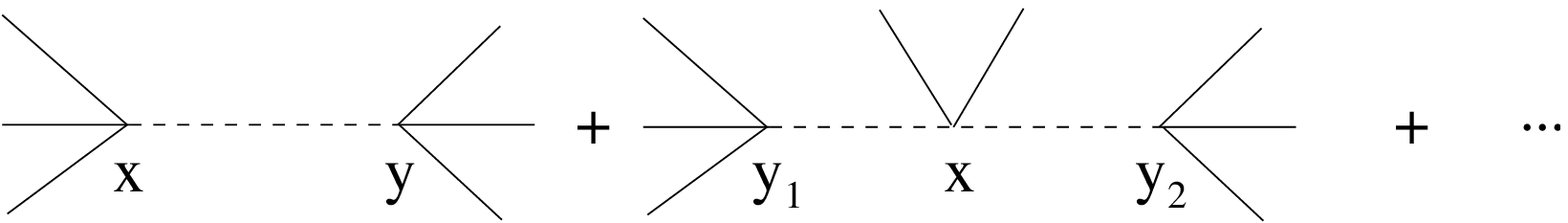}
\caption{Perturbative expansion of the zero-mode induced parts of the
interaction Hamiltonian $H_{int}^{(0)}(\varphi)$. The ``external
lines'' are $p^+=0$ combinations of operators, $[\varphi^2]_0$ or
$[\varphi^3]_0$.}
\end{minipage}
\end{center}
\end{figure}

Plugging the solution $\phi_0^{(0)}=\phi_0^{(0)}(\varphi)$ into the
original Hamiltonian, $\mathcal{H}= \frac{1}{2}(\varphi
+\phi_0^{(0)})(m^2) (\varphi +\phi_0^{(0)}) + \frac{\lambda}{4!}
(\varphi +\phi_0^{(0)})^4=\frac{1}{2} m^2 \varphi^2 +
\mathcal{H}_{int}^{(0)}(\varphi) $, we obtain the effective
interaction Hamiltonian $\mathcal{H}_{int}^{(0)}(\varphi)$ whose
perturbative expansion reads
\cite{M&R}:
\begin{eqnarray}
{}-i\int d x^+ H_{int}^{(0)}(\varphi)
  &=& -i\lambda\int d^2 x \frac{1}{4!} \varphi^4(x)
	+\frac{1}{2!}(-i\lambda)^2 \int d^2 x \int d^2 y
	\left(\frac{\varphi^3(x)}{3!} \Delta_0(x-y)
	\frac{\varphi^3(y)}{3!} \right)\nonumber \\
  &+&\frac{3}{3!}(-i\lambda)^3 \int d^2 x \int d^2 y \int d^2 z
	\left(\frac{\varphi^3(y)}{3!} \Delta_0(y-x)
	\frac{\varphi^2(x)}{2!} \Delta_0(x-z) \frac{\varphi^3(z)}{3!}
	\right) \nonumber \\
  && +~O(\lambda^4)\, ,
\label{effint}
\end{eqnarray}
which has the zero-mode-induced terms (see Fig. 2) in addition to the
original term (first term). Note again that zero-mode appears only in
the tree internal line coupled to $[\varphi^2]_0$ and
$[\varphi^3]_0$.

Now, the perturbative S matrix is given by the expansion of
$S=T\exp{\left[-i\int_{-\infty}^{\infty}dx^{+}
H_{int}^{(0)}(\varphi)\right]}$.
It yields Feynman graphs which contain no closed single line of the
zero-mode propagator, since there is no zero mode operator to be
contracted in $H_{int}(\varphi)$, namely {\it the zero-mode loop is
absent}.

\section{NON-COVARIANT RESULT}
Let us now show that as a simplest example the two-body scattering
amplitude (forward scattering amplitude) with zero $p^+$ transfer in
DLCQ disagrees with the covariant one \cite{TUYY}.
At one loop the covariant amplitude in this case is given by
\begin{eqnarray}
  A_{\rm Cov}^{\rm Forward} &=& \frac{1}{2}(-\lambda)^2 \int
	\frac{dk^+dk^-}{(2\pi)^2i}\left(
	\frac{1}{m^2 - 2k^+k^-}\right)^2 \nonumber\\
  &=&\frac{1}{2}(-\lambda)^2\int_{0}^{\infty}\frac{dk^+}{4\pi(k^+)^2}
	\int_0^{\infty}d\xi \,\xi\,\delta(\xi)\,
	\exp\left(-\frac{\xi m^2}{2k^+}\right)\
	=\frac{(-\lambda)^2}{8\pi m^2} \ne 0 \,,
\label{Covariant}
\end{eqnarray}
which is {\it non-vanishing} result, where the integral $d k^+$ should
be done before that of $d \xi$, since otherwise $d \xi$ first would
give $0$ but $d k^+$ afterward would divergence and hence the integral
would be ill-defined $\infty \times 0$.
On the other hand, straightforward calculation via the DLCQ
perturbation mentioned above yields for the same process the {\it
vanishing} amplitude:
\begin{eqnarray}
 A_{\rm DLCQ}^{\rm Forward} = \frac{1}{2} (-\lambda)^2
	\frac{1}{2L}\sum_{l>0}\frac{1}{2(\frac{l\pi}{L})^2}
	\int_0^{\infty}d\xi\,\xi\,\delta(\xi)\,
	\exp\left(-\frac{\xi m^2}{2 (l\pi/L)}\right)=0\,
\end{eqnarray}
{\it independently of} $L$ and so does  in the continuum limit
$L \rightarrow \infty$.
Note that the DLCQ amplitude differs from (\ref{Covariant}) simply by
the replacement
$1/(2\pi)\int_0^{\infty}dk^+\rightarrow 1/(2L)\sum_{l>0}\, ,$
where {\it the zero mode loop is absent} $l\ne 0$ in the DLCQ
amplitude, and this time the integral $d \xi$ can be done before $
\sum_{l>0}$, since $(2L)^{-1}\sum_{l>0}\,[2 (l\pi/L)^2]^{-1}
= L/24 < \infty$ and hence the calculation is well-defined,
$({\rm finite}) \times 0=0$.

The discrepancy is just the forward scattering amplitude with measure
zero but has serious physical consequences. Actually, the same
arguments can apply to the loop diagrams attached with arbitrary
number of sets of external lines with $p^+=0$ corresponding to
$[\varphi^2]_0$ and $[\varphi^3]_0$.
Vanishing of all these types of diagrams implies that the effective
potential is zero, for example, as was emphasized in \cite{Heinzl}.
 
The absence of the zero-mode loop can be seen more transparently in the
path integral formalism \cite{TUYY2}
where the second-class constraints $\Phi_1\equiv
\pi_{\varphi}-\partial_- \varphi, \Phi_2\equiv \pi_0, \Phi_3\equiv
[\frac{\partial \mathcal{H}}{\partial \phi}]_0$ are incorporated in
the standard way:
\begin{eqnarray}
Z&=&\int \left[D\pi_{\varphi} D\pi_0 D\varphi D\phi_0\right]
  \delta(\Phi_1)\delta(\Phi_2) \delta(\Phi_3) (Det\, C)^{1/2}\,
  \exp \left[i \int \left(\pi \partial_+\phi-\mathcal{H}\right)\right]
	\nonumber\\
 &=& \int \left[D\varphi D\phi_0 D B Dc D{\bar c}\right]\,
    \exp\left[ i \int \left\{\mathcal{L} +\left(B \left[
    \frac{\partial \mathcal{H}}{\partial \phi}\right]_0
    +i {\bar c} \left[\frac{\partial^2 \mathcal{H}}{\partial\phi^2}
	\right]_0 c\right)\right\}\right]\, ,
\end{eqnarray}
where the integral
$D\pi_{\varphi} D\pi_0 \delta(\Phi_1)\delta(\Phi_2)$ was done
trivially, and $\delta(\Phi_3) = \int [D B]
\exp[i\int B [\frac{\partial \mathcal{H}}{\partial \phi}]_0] $
and $(Det \, C)^{1/2} = Det \left(\left[\frac{\partial^2
\mathcal{H}}{\partial \phi^2} \right]_0\right) =\int [D c D \bar c]
\exp \left[i \int i {\bar c} \left[\frac{\partial^2
\mathcal{H}}{\partial \phi^2}\right]_0  c\right]$, with $B,c, {\bar
c}$ being Nakanishi-Lautrup field, ghost and antighost,
respectively.\footnote{If we perform the integral of the zero mode $[D
\phi_0] \delta(\Phi_3) (Det \, C)^{1/2}$, we obtain
$Z=\int [D\varphi] \exp \left[i\int\mathcal{L}^{(0)}(\varphi)\right]$,
where $ \mathcal{L}^{(0)}(\varphi)$ is the effective Lagrangian
written only in terms of non-zero modes, corresponding to
(\ref{effint}), which yields the same Feynman rule as in section
2. For the SSB solution $\phi_0^{(\pm v)}$, we make a simple
replacement: $\mathcal{L}^{(0)}(\varphi) \rightarrow \mathcal{L}^{(\pm
v)}(\varphi)$.}
 
Now, $\left(B\left[\frac{\partial\mathcal{H}}{\partial\phi}\right]_0
+i{\bar c}\left[\frac{\partial^2 \mathcal{H}}{\partial\phi^2}\right]_0
 c \right)=\delta_{\rm BRS} \left(-i {\bar c} \left[
\frac{\partial \mathcal{H}}{\partial \phi}\right]_0 \right)$ is a BRS
singlet, where the BRS transformation $\delta_{\rm BRS}$ is defined as
$\delta_{\rm BRS} {\bar c} =iB, \delta_{\rm BRS} \phi_0 =c,
\delta_{\rm BRS} B=\delta_{\rm BRS} c =0$.
Then $(\phi_0,B,{\bar c},c)$ {\it are BRS-quartet and hence the loop
of $(\phi_0, B, {\bar c}, c)$ cancel each other} in much the same way
as in the gauge theories \cite{K&O}.
We have explicitly checked this cancellation of the zero-mode loop by
the $B, {\bar c},c$ loops to two-loop order. In contrast, the covariant
expression has no constraint and simply  $Z=\int[D\phi]
\exp\left[ i \int \mathcal{L}\right]$ where the loop effects
of the zero-mode (though not clearly separated) are not canceled out.

\section{RECOVERING LORENTZ INVARIANCE}
We have seen that the violation of Lorentz invariance is a real effect
in DLCQ as far as we use the naive discretization
$\int_{-\infty}^{+\infty} d x^-\mathcal{L} \rightarrow \int_{-L}^{+L}
d x^-\mathcal{L}$, with $\mathcal{L}$ in DLCQ being {\it the same} as
that of the continuum theory. However we know in the  lattice theory
that the discretized action should be different in principle from
that of the continuum theory: Lorentz covariant theory should be
constructed by {\it using the solution} which has the second order
phase transition and hence yields the sensible continuum limit.
 
In much the same spirit, here we propose the new DLCQ Lagrangian
modified by adding extra operators $-\Delta H$ in such a way that
{\it the perturbative solution} yields the covariant limit
\cite{TUYY2}. The extra term is generated by the zero-mode loop
which can be explicitly estimated by the covariant perturbation
theory. Important point of our method is that the {\it trivial vacuum
is not destroyed} by the extra term.
 
At one loop order of the zero mode loop we have found the compact form
of the extra operators in the Hamiltonian $\Delta H$:
\begin{eqnarray}
 \frac{\Delta H}{2L}&=&\frac{1}{2}\int\frac{d k^+ d k^-}{(2\pi)^2i}
    \log\left[\frac{m^2+\frac{\lambda}{2!}\left(
    [\varphi^2]_0+(\phi_0^{(0, \pm v)})^2\right)-k^+ k^-}
    {m^2-k^+k^-}\right] \nonumber \\
 &=&-\sum_{n=1}^{\infty}
  \left(-\frac{\lambda}{2!}\right)^n\frac{1}{2n}I_n
  \left([\varphi^2]_0+(\phi_0^{(0,\pm v)})^2\right)^n\,,\label{deltah}
\end{eqnarray}
\begin{figure}
\begin{center}
\includegraphics[width=35mm]{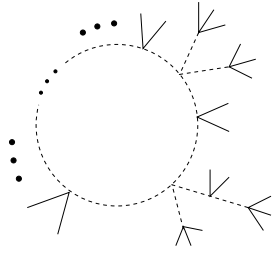}
\caption{Perturbative amplitude induced by the zero-mode loop.}
\end{center}
\end{figure}
where $\phi_0^{(0, \pm v)}$ is an operator solution of ZMC in
(\ref{Sol}) and $I_n (n\geq1)
=\int\frac{dk^+dk^-}{(2\pi)^2i}(m^2-k^+k^-)^{-n}$.
($n=1$, the $\log$ divergence is regularized by Pauli-Villars
regularization).
This yields the correct Lorentz invariant scattering amplitude for
any number of external legs with zero $p^+$ transfer, i.e.,
$[\varphi^2]_0$ and/or $[\varphi^3]_0$ (see Fig. 3).
The term $n=2$, for example, indeed yields the Lorentz-invariant
two-body forward scattering amplitude with zero $p^+$ exchange in
(\ref{Covariant}).
It is obvious that the extra term $\Delta H$ yields the correct
effective potential by simply replacing  $[\varphi^2]_0+(\phi_0^{(0,
\pm v)})^2$ by a c-number constant $\phi_c^2$.

\section{CONCLUSION}

We have shown that the DLCQ with naive discretization violates Lorentz
invariance in the perturbative S matrix due to the lack of zero mode
loop, while DLCQ guarantees that the trivial vacuum is a true vacuum
even when the SSB takes place.
We have proposed to modify the DLCQ action by adding extra operators
in such a way that the continuum limit S matrix recovers Lorentz
invariance without spoiling the trivial vacuum. We have given such a
modification explicitly at one loop level.
Study at higher loops is in progress. Similar arguments will be
applied to four dimensional theories but meet with substantial
complexity due to the transverse degrees of freedom. This is also
under investigation.
Although our arguments are only within perturbation, we expect that
our strategy to modify the DLCQ action based on the solution of the
dynamics should be the right way to recover the Lorentz invariance
without spoiling the trivial vacuum even in the non-perturbative
approach.


\section*{ACKNOWLEDGEMENTS}
This work is supported in part by JSPS Grant-in-Aid for the
Scientific Research (B)(2) \#14340072 (S.Y. and K.Y.) and MEXT
Grant-in-Aid for the Scientific Research \#13135212 (S.U.).

\end{document}